\documentclass{statsoc}

\advance\hoffset by -1in
\advance\voffset by -1in

\usepackage{csquotes}

\usepackage{amsmath,amssymb,mdwlist}

\usepackage{ifthen}
\usepackage[usenames]{color}
\newcounter{content}
\newcounter{notes}
\newcounter{techreport}
\newcounter{app}

\newcommand{\content}[1]{\ifthenelse{\equal{\value{content}}{0}}{}{#1}}
\newcommand{\app}[1]{\ifthenelse{\equal{\value{app}}{0}}{}{#1}}
\newcommand{\TRonly}[1]{\ifthenelse{\equal{\value{techreport}}{0}}{}{#1}}
\newcommand{\notes}[1]{\ifthenelse{\equal{\value{notes}}{0}}{}{{\color{blue} \textbf{Note: }{#1}\\}}}
\newcommand{\ignore}[1]{}
\newcommand{\needcite}{\ifthenelse{\equal{\value{notes}}{0}}{}{{\color{red} \citep{citation}}}}

\usepackage{mathtools}
\usepackage{todonotes} 
\usepackage{verbatim}
\usepackage{amsfonts}
\usepackage{csquotes}
\usepackage[noend]{algorithmic}
\usepackage{multirow}
\usepackage{palatino, url, multicol}
\newcounter{blind}
\newcommand{\blind}[2]{\ifthenelse{\equal{\value{blind}}{0}}{#1}{#2}}
\usepackage{ifthen}
\newboolean{draft}
\setboolean{draft}{true}  
\newcommand{\inote}[1]{\ifthenelse{\boolean{draft}}{{\bf IAN:~}{\it
#1}\relax}{}}
\newcommand{\mnote}[1]{\ifthenelse{\boolean{draft}}{{\bf MARK:~}{\it
#1}\relax}{}}

\newcommand{\Perp}{\perp \! \! \! \perp}

\title{Exponential-family Random Network Models}
\author[Fellows and Handcock]{Ian Fellows and Mark S. Handcock}
\email{ian.fellows@stat.ucla.edu}
\address{University of California, Los Angeles, CA, USA} \coaddress{Ian
Fellows,
Department of Statistics, University of California, Los Angeles, CA 90095-1554,
U.S.A.}
\begin{document}
\maketitle

\begin{abstract}
Random graphs, where the connections between nodes are considered random variables, have wide applicability in the social sciences. Exponential-family Random Graph Models (ERGM) have shown themselves to be a useful class of models for representing complex social phenomena. We generalize ERGM by also modeling nodal attributes as random variates, thus creating a random model of the full network, which we call Exponential-family Random Network Models (ERNM).
We demonstrate how this framework allows a new formulation for logistic 
regression in network data.
We develop likelihood-based inference for the model and an MCMC algorithm to
implement it.

This new model formulation is used to analyze a peer social network from the National Longitudinal Study of Adolescent Health.
We model the relationship between substance use and friendship relations, and
show how the results differ from the standard use of logistic regression on
network data.

%
\end{abstract}
\blind{}{
\vspace*{.3in}
}

\advance\abovedisplayskip by -5pt
\advance\belowdisplayskip by -5pt

\section{Introduction}

Random graphs, where connections between nodes are random but nodal characteristics are either fixed or missing, have a long history in the mathematical literature starting with the simple Erd\H{o}s-R\'{e}nyi model \citep{ErdosRennyi_1959}, and including the more general  exponential-family random graph models (ERGM) for which inference requires modern Markov Chain Monte Carlo (MCMC) methods \citep{fra86,hunhan04}. On the other hand we have Gibbs/Markov random field models where nodal attributes are random but interconnections between nodes are fixed. A simple example is the Ising model of ferromagnetism \citep{Ising_1925} from the statistical physics literature which is exactly solvable under certain network configurations \citep{Baxter_1982}; however, most field models require more complex methodologies for inference \citep{ZhuLiu_2002}.

In the social network literature, these two classes of models are conceptually defined as ``social selection'' and ``social influence'' models. In social selection models, the probability of social ties between individuals are determined by nodal characteristics such as age or sex (see \citet{RobinsElliottPattison_2001} and references therein). In social influence models, individuals' nodal characteristics are determined by social ties (see \citet{RobinsPattisonElliott_2001} and references therein). \citet{Leenders_1997} argues that the processes of tie selection and nodal variate influence are co-occurring phenomena, with ties affecting nodal variates and visa versa, and should therefore be considered together. This paper presents  a joint exponential-family model of connections between nodes (dyads), and nodal attributes, thus representing a unification of social selection and influence. We will refer to this model as an exponential-family random network model (ERNM).


We note that we are not developing a model for the coevolution of the tie and nodal variables. We are modeling the joint relation between the processes of tie selection and nodal variate influence in a cross-sectional network. As such our model explicitly represents the endogenous nature of the relational ties and nodal variables. If network-behavior panel data is available then it may be possible to statistically  separate  the effects of selection from those of influence. For a discussion of these issues for dynamic and longitudinal data, see \citet{steglich2010}.

The next section (Section \ref{sec:spec}) introduces the ERNM class and gives simple examples.
Section \ref{sec:dev} develops aspects of the class that are important for
statistical modeling. 
Section \ref{sec:app} applies the modeling approach to the study of substance abuse in
adolescent peer networks and compares it to standard approaches.
Section \ref{sec:diss} concludes the paper with a broader discussion.

\section{ERNM specification}\label{sec:spec}

Let $Y$ be an $n$ by $n$ matrix whose entries $Y_{i,j}$ indicate whether subject $i$ and $j$ are connected, where $n$ is the size of the population. Further let $X$ be an $n\times q$ matrix of nodal variates. We define the {\it network} to be the random variable $(Y,X)$. 
Let ${\cal N}$ be the set of possible networks of interest (the sample space of the model). For example, ${\cal N} \subseteq 2^{\mathbb Y} \times {\cal X}^n$, the power set of the dyads in the network times the power set of the sample space of the nodal variates. 
A joint exponential family model for the network may be written as:

\begin{equation} \label{eq:lik}
P(X=x,Y=y | \eta) = \frac{1}{c(\eta,{\cal N})}e^{\eta{\cdot}g(y,x)}, ~~~~~~(y,x)\in {\cal N}
\end{equation}

{\noindent}where $\eta$ is a vector of parameters, $g$ is a vector valued function, and $c(\eta,{\cal N})$ is a normalizing constant such that the integral of $P$ over the sample space of $X$ and $Y$ is 1 (See equation \eqref{eq:norm}). The model parameter space is $\eta\in H\subseteq{\mathbb R}^q$. This functional form is the familiar exponential family form, and is extremely general depending on the choice of $g$ (see \citet{bar78} and \citet{krivitskyvalued11}).
Formally, let $({\it N}, {\cal N}, P_0)$ 
be a $\sigma-$finite measure space with
reference measure $P_0.$ A probability measure $P(X=x,Y=y | \eta)$ is an ERNM with
respect to this space if it
is dominated by $P_0$ and the Radon-Nikodym derivative of $P(X=x,Y=y | \eta)$ with
respect to $P_0$ is expressible as:
\begin{equation} \label{eq:rnd}
\frac{dP(X=x,Y=y | \eta)}{dP_0} = \frac{1}{c(\eta,{\cal N})}e^{\eta{\cdot}g(y,x)}, ~~~~~~(y,x)\in {\cal N}\nonumber
\end{equation}

{\noindent}where

\begin{equation} \label{eq:norm}
c(\eta,{\cal N}) = \int_{(y,x)\in\cal N}{e^{\eta{\cdot}g(y,x)}dP_0(y,x)}
\end{equation}

{\noindent}and $H \subseteq \{\eta\in{\mathbb R}^q : c(\eta,{\cal N}) < \infty\}$.
See \citet{bar78} for further properties of the exponential-family class of probability distributions.


\subsection{Relationship with ERGM and Random Fields}\label{ernm:relationship}

Let ${\cal N}(x)= \{y: (y,x)\in {\cal N}\}$ and
 ${\cal N}(y)= \{y: (y,x)\in {\cal N}\}$ then
\begin{eqnarray}
P(Y=y | X=x; \eta) &=& \frac{1}{c(\eta;{\cal N}(x),x)}e^{\eta {\cdot}g(y,x)}~~~y \in {\cal N}(x) \nonumber \\
P(X=x | Y=y; \eta) &=& \frac{1}{c(\eta;{\cal N}(y),y)}e^{\eta {\cdot}g(y,x)}~~~x \in {\cal N}(y) \nonumber 
\end{eqnarray}
The first model is the ERGM for the network conditional on the
nodal attributes.  Analysis of models of this kind have been the
staple of ERGM \citep{fra86,hunhan04, goodkittsmorris09}.
The second model is an exponential-family for the field of nodal attributes
conditional on the network. 
This will be a Gibbs/Markov field when the process satisfies the
pairwise Markov property (i.e., If $Y_{ij}=0$ then $X_i$ and $X_j$ are
conditionally independent given all other $X$) \citep{bes74}. However the model is more
general than this as $g(y,x)$ can be arbitrary. We will refer to it as a Gibbs
measure \citep{geo88}.



The model \eqref{eq:lik} can be expressed as
\begin{equation}
P(X=x,Y=y | \eta) = P(Y=y | X=x | \eta) P(X=x | \eta)
\end{equation}
where
\begin{equation*}
P(X=x| \eta) = \frac{c(\eta;{\cal N}(x),x)}{c(\eta,{\cal N})}~~~x \in {\cal X}
\end{equation*}
This model is the marginal representation of the nodal attributes and is not
necessarily an exponential-family with canonical parameter $\eta.$ These
decompositions demonstrate why the 
joint modeling of $Y$ and $X$ via ERNM (as proposed here) is different and 
novel compared to the conditional modeling of $Y$ given $X$ via
ERGM.

\subsection{Interesting model-classes of ERNM}\label{ernm:interesting}

\subsubsection{Example: Separable ERGM and Field Models}\label{ss:separable}

Suppose that $g$ is composed such that the model can be expressed as
\begin{equation}\label{separable}
P(X=x,Y=y | \eta_1,\eta_2) = \frac{1}{c(\eta_1, \eta_2,{\cal N})}e^{\eta_1{\cdot}h(x) + \eta_2{\cdot}g(y)}~~~~~(y,x)\in {\cal N}.
\end{equation}
where ${\cal N}$ is the product space ${\cal Y}\times{\cal X}$ with ${\cal Y}$ pertaining to $Y$ and ${\cal X}$ to $X$. $x$ and $y$ in this model are  separable and therefore may be considered independently. The model \eqref{separable} can be decomposed as the product of
\begin{eqnarray}
P(X=x | \eta_1) &=& \frac{1}{c_1(\eta_1,{\cal X})}e^{\eta_1 {\cdot}h(x)} \nonumber \\
P(Y=y | \eta_2) &=& \frac{1}{c_2(\eta_2,{\cal Y})}e^{\eta_2 {\cdot}g(y)}. \nonumber
\end{eqnarray}
This type of model is particularly simple because of the separation of the two components. The first term is a general exponential-family model for the attributes (e.g., generalized linear models \citet{mccullaghnelder1989}). The second term is a separate ERGM for the relations that has no dependence on the nodal attributes. Such separable models are usually not applicable as the phenomena that we are interested in studying is precisely the relationship between $X$ and $Y$, thus independence is typically an unrealistic assumption.

\subsubsection{Example: Joint Ising Models}\label{ss:ising}

If $X$ is univariate and binary $x_i\in {\{-1,1\}}$, previous social selection models \citep{goodkittsmorris09} have used the following statistic to model homophily
\begin{equation}\label{eq:homophily}
{\rm homophily}(y,x) = \sum_{i=1}^n\sum_{j=1}^n x_iy_{i,j}x_j
\end{equation}
It counts the number of ties between nodes homophilous in the nodal covariate. Such a statistic is useful as a basis for a joint model. A simple example would include a term for homophily and a term graph density, explicitly
$$
P(X=x,Y=y | \eta_1, \eta_2) \propto e^{\eta_1 {\rm density}(y) + \eta_2{\rm homophily}(y,x)}~~~~~~
(y,x) \in {\cal N}.
$$
{\noindent}where {\rm density}(y) = $\frac{1}{n}\sum_i\sum_jy_{i,j}$
and ${\cal N} = {\cal Y}\times{\cal X} = \{0,1\}^{2^n} \times \{-1,1\}^n$.
If we look at the conditional distribution of $Y$ given $X$ we get
$$
P(Y_{i,j}=y_{i,j} | X=x, \eta_1,\eta_2) \propto e^{\eta_1 \frac{1}{n}y_{i,j} + \eta_2x_iy_{i,j}x_j}~~~y\in\{0,1\},~x\in{\cal X}.
$$
Note that the dyadic variables $y_{i,j}$ are independent of each other, so that this is a so called dyad-independent model for $Y$. We can recognize the functional form of the conditional distribution of $Y$ given $X$ as identical to logistic regression, and thus the conditional likelihood could be maximized using familiar generalized linear model (GLM) algorithms \citep{mccullaghnelder1989}. Conditioning $X$ on $Y$ we arrive at
$$
P(X=x| Y=y, \eta_2) \propto e^{\eta_2\sum_i\sum_j x_iy_{i,j}x_j}~~~~~
(y,x) \in {\cal N},
$$
which we can recognize as the familiar Ising model \citep{Ising_1925} for the field over $X$ with its lattice defined by $Y$. 

This joint Ising model has the advantage of being mathematically parsimonious. Unfortunately, the results in section \ref{ss:deg} indicate that it displays unrealistic statistical characteristics, which may rule it out as a reasonable representation of typical social networks.

\section{Development of ERNM}\label{sec:dev}

In this section we develop ERNM, including issues of model
degeneracy, the specification of network statistics and likelihood-based
inference. In particular, we specify a class of logistic regression models for
ERNM that represent the endogeneity of the nodal attributes.

A large component of modeling with the ERNM class is the specification of 
the statistics $g(y,x)$. As each choice of $g(y,x)$ leads to a valid model for
the network process, there is much flexibility in this for modeling. 
The particular choices are very application dependent. However, as for ERGM, a
stable of statistics can be created to capture primary features of networks
such as density, mutuality of ties, homophily, reciprocity,
individual heterogeneity in the propensity to form ties,
and the transitivity of relationships between actors \citep{ergmtermsjss}.

It is important to note that the ERNM class is quite different from the ERGM
class (despite the formal similarity in equation (1)). ERNM require the
specification of stochastic models for the nodal attributes (which ERGM do not
permit). Further statistics which are meaningless for ERGM, for example, any
statistic of $X$ alone, play a prominent role in ERNM.

\subsection{Model Degeneracy}\label{ss:deg}

Exponential family models for networks have been known to suffer from model degeneracy \citep{str86,handcock2002nas,schweinberger2011}, and even simple Markov models have similarly been shown to have degenerate states (sometimes called phase transitions in the statistical physics literature \citep{dyson69}). Because ERNM models represent the unification of these two classes of models, a consideration of degeneracy must be undertaken. For example, while the joint Ising model of Section \ref{ss:ising} is pleasing in its parsimonious simplicity, it unfortunately displays pathological degeneracy under mild homophily conditions. Consider a 20 node network, with $\eta_1=0$ and $\eta_2=0.13$. In this model, 76\% of edges are between nodes with matching $x$ values, whereas 24\% are between miss-matched nodes. Figure 1 shows the marginal statistics of 100,000 draws from this model.

\begin{figure}
\centering
\includegraphics[scale=0.6]{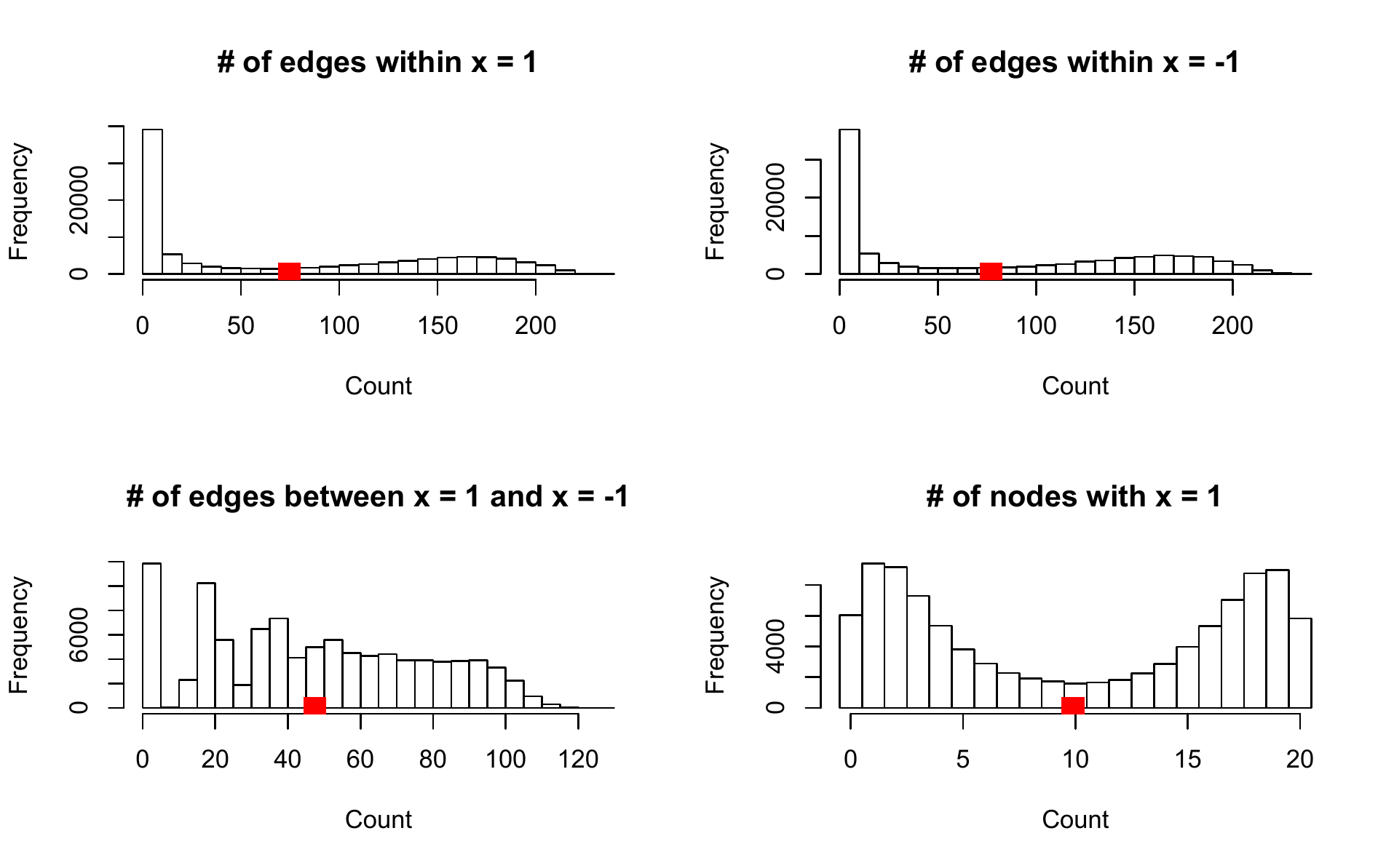}
\caption{\label{figure:f1}
100,000 draws from an Ising Joint Model with $\eta_1=0$ and $\eta_2=0.13$. Mean values are marked in red.}
\end{figure}

Despite the fact that the homophily is not particularly severe, Figure 1 displays a great deal of degeneracy. The counts of edges are highly skewed. By symmetry we know that the expected number of nodes with $x = 1$ is 10, however, when inspecting the marginal histogram, we see that it is bimodal and puts very low probability on the value of 10. This severe degeneracy greatly reduces the usefulness of this model for practical networks.

We note that this phenomena will likely be as prevalent for ERNM models as for ERGM, and will have similar solutions. We recommend that model degeneracy be assessed for all proposed ERNM models.

\subsection{Non-degenerate representation of Homophily within ERNM}\label{ss:homo}

Specification of the network's statistics via $g$ is fundamental to ERNM.
A natural source are analogues of those terms developed for ERGM \citep{ergmtermsjss}.
However, the degeneracy of the homophily specification in Section \ref{ss:ising} suggests
that careful thought is required in considering some network statistics.  
Suppose $x$ is categorical with category labels $1,\ldots,K$.
To define homophily we start by defining fundamental statistics of the
network.
Let $d_{i}(y)$ be the degree of node $i=1,\ldots,n$ 
and $n_k(x) = \sum_{i}{I(x_i=k)}$
be the category counts, that is, 
the number of nodes in category $k=1,\ldots, K$. Here $I$ is the indicator function.
Let $d_{i,k}(y,x) = \sum_{i<j}{y_{ij}I(x_j=k)}$ be
the number of edges connecting node $i$ to nodes in category $k$. 
We can generalize Equation \eqref{eq:homophily} as
$$
{\rm homophily}_{k,l}(y,x) = \sum_{i=1}^n\sum_{j=1}^n I(x_i=k)y_{i,j}I(x_j=l).
$$
As with Equation \eqref{eq:homophily}, this term has the nice property that it is dyad independent, meaning that conditional upon $X$, the marginal distribution of each dyad is independent of all others. Unfortunately, it displays the same degeneracy we saw in Section \ref{ss:ising}. We propose an alternate regularized homophily statistic which can be expressed as
$$
{\rm rhomophily}_{k,l}(y,x) = \sum_{i:x_i=k} \sqrt{d_{i,l}(y,x)} - E_{\Perp}(\sqrt{d_{i,l}(Y,X)} | Y=y, n(X)=n(x)),
$$
where $E_{\Perp}(g(Y,X) | Y=y, n(X)=n(x))$ is the expectation of the 
statistic $g(Y,X)$ conditional upon the graph $Y$ and number of nodes in each category of $x$ ($n(x) = \{n_k(x)\}_{k=1}^{K}$), under the assumption that $X$ and $Y$ are independent. Specifically, this distribution is
$$
P(X=x | Y=y, n(X)=n(x))
\propto 1
~~~~~~~~~~~~~~~~~~~~~~~ (y,x) \in {\cal N},
$$

There are many possible definitions of homophily, and this is one of many ways to formulate the relationship and in some applications, there may be a superior form. The justification for this particular formula is primarily empirical in that it captures the relationship between nodal variates and dyads well, and does not display the degeneracy issues that plague other forms of homophily. There are, however, some features of the statistic which provide justification for its form. The statistic $d_{i,l}(y,x)$ is transformed by a square root to roughly stabilize the variance based on the Poisson count model. This is important as nodes with high degree should not have qualitatively larger influence than nodes with low degree. Subtracting off the expectation based on the uniform independence model is essential in avoiding degeneracy because degenerate networks where all, or almost all, nodes belong to the same category should have homophily near zero.

\subsection{Logistic Regression for Network Data}

Let us consider a specific form of Equation \eqref{eq:lik} were $X$ is partitioned into a binary nodal variate of particular interest $Z\in \{0,1\}$ (i.e. an outcome variable), and a matrix of regressors $X$.

\begin{equation}
P(Z=z, X=x,Y=y | \eta, \beta, \lambda) = \frac{1}{c(\beta,\eta,\lambda)}e^{z{\cdot}x\beta + \eta{\cdot}g(y,x) + \lambda{\cdot}h(y,z) }.
\end{equation}

We can then write the distribution of $z_i$ conditional upon all other variables as

\begin{equation} \label{eq:lrcond}
P(z_i=1 | z_{-i}, x_i,Y=y, \beta, \lambda) = \frac{ e^{x_i\beta } }{ e^{\lambda{\cdot}[h(y,z^-)-h(y,z^+)] }  + e^{x_i\beta }  }.
\end{equation}
where $z_{-i}$ represents the set of $z$ not including $z_i$, $z^+$ represents the variant of $z$ where $z_i=1$, $z^-$ is the variant of $z$ where $z_i=0$, and $x_i$ represents the $i$th row of $X$. Suppose all variables remain fixed at their value except for $x_i$, which changes to $x_i^*$, then using equation \eqref{eq:lrcond}, we can write the log odds ratio as
$$
{\rm logodds}(z_i=1 | z_{-i}, x_i,Y=y, \beta, \lambda) - {\rm logodds}(z_i=1 | z_{-i}, x_i^*,Y=y, \beta, \lambda) = \beta(x_i - x_i^*).
$$
Thus, the coefficients $\beta$ may be interpreted as a conditional logistic regression model (i.e. conditional upon the rest of the network, a unit change in $x_i$ leads to a $\beta$ change in the log odds). Though the interpretation of the coefficients is familiar, the usual algorithms for estimating a logistic regression can not be used because the distribution of $z_i$ depends on $z_{-i}$ and thus the independence assumption does not hold.

\subsection{Likelihood-based Inference for ERNM}\label{ss:inference}

The likelihood in equation \eqref{eq:lik} can be maximized using the methods of \citet{GeyerThompson_1992} and \citet{hunhan04}. Let $y_{obs}$ and $x_{obs}$ be the observed network, and $\ell$ be the log likelihood function. The log likelihood ratio for parameter $\eta$ relative to $\eta_0$ can be written as,
\begin{equation*}
\ell(\eta) - \ell(\eta_0) = (\eta - \eta_0){\cdot} g(y_{obs}, y_{obs}) - \log[E_{\eta_0}(e^{(\eta - \eta_0){\cdot} g(y,x)}])
\end{equation*}

Given a sample of $m$ networks ($y_i$, $x_i$) from $P(X=x,Y=y | \eta_0)$ the log likelihood can be approximated by
\begin{equation} \label{eq:likapprox}
\ell(\eta) - \ell(\eta_0) \approx (\eta - \eta_0){\cdot} g(y_{obs}, x_{obs}) - \log(\frac{1}{m}\sum_{i=1}^me^{(\eta - \eta_0){\cdot}g(y_i,x_i)}))
\end{equation}

Appendix B provides the details of the Metropolis-Hastings algorithm used to sample from $P(X=x,Y=y | \eta_0)$ when the normalizing constant $c$ is intractable (which is usually the case). The approximation in equation \eqref{eq:likapprox} degrades as $\eta$ diverges from $\eta_0$, motivating the following algorithm for estimating the maximum likelihood parameter estimates

\begin{enumerate}
\item Choose initial parameter values $\eta_0$.
\item Use Markov Chain Monte Carlo to generate $m$ samples $(y_i, x_i)$ from $P(X=x, Y=y | \eta_0 )$.
\item With the sample from step 2, find $\eta_1$ maximizing a H\"ajek estimator \citep{thompson2002} of Equation \eqref{eq:likapprox} subject to ${\rm abs}(\eta_1 - \eta_0) < \epsilon$.
\item If convergence is not met, let $\eta_0 = \eta_1$ and go to step 2.
\end{enumerate}

This approximation to the log-likelihood can then be used to derive the
Fisher information matrix and other quantities used for inference. Note that the usual
asymptotic approximations based on $n\to\infty$ may not apply to this situation
as $n$ is often endogenous to the social process.

\section{Application to substance use in adolescent peer networks}\label{sec:app}

In addition to collecting data on the health related behaviors, the National Longitudinal Study of Adolescent Health (Add Health) also collected information on the social networks of the subjects studied \citep{Harris_2003}. 

The network data we study in this article was collected during
the first wave of the study. The Add Health data came from a stratified
sample of schools in the US containing students in grades 7
through 12; the first wave was conducted in 1994-1995. For the
friendship networks data, Add Health staff constructed a roster of
all students in the school from school administrators.  Students
were then provided with the roster and asked to select up to five
close male friends and five close female friends.
Complete details of this and subsequent waves of the study
can be found in \citet{resnick1997} and \citet{udrybear98}.


Previous studies have investigated the social network structure of Add Health schools \citep{bearman.et.al:ajs:2004}, including \citet{hungoodhan07,goodkittsmorris09,hangile07} who used ERGM models to investigate network structure.

Here we analyze one of these schools; the high school had 98 students, of which 74 completed surveys. Students who did not complete the survey were excluded from analysis. The data contains many measurements on
each of the individuals in these networks with some
measurements, like sex, not influenced by network structure in
any way,  termed {\em exogenous}. Other covariates may
exhibit strong non-exogeneity (e.g., substance use may be influenced
through friendships).  

\subsection{A Super-population Model for an Add Health High School}\label{ss:super}

Using the MCMC-MLE algorithm in Section \ref{ss:inference}, we fit an ERNM model to the high school data. The model has six terms modeling the degree structure of the network, three modeling the counts of students in each grade, and two representing the homophily within and between grades. Table \ref{tbl:modterms} defines each of the terms, and explicit formulas are listed in Appendix A. Note that many terms could be added to this model to make it a more complex representation of the social structure, including terms similar to those in \citet{hangile07}, however, here we prefer a simple parsimonious model of the network, with particular focus on the relationship between $X$ and $Y$.

\begin{table}
  \caption{ERNM Model Terms: The terms in the first block are graph statistics (ERGM-type), those in the second block model nodal attributes, and the last are joint. Terms in the the last two blocks can not be represented in an ERGM.}
   \centering
\begin{tabular}{lrl}
  \hline
Form & Name & Definition \\ 
  \hline
$Y$ & Mean Degree & Average degree of students \\ 
$Y$ &  Log Variance of Degree & The log of the variance of the student degrees \\ 
$Y$ &  In Degree = 0 & \# of students with in degree 0 \\ 
$Y$ &  In Degree = 1 & \# of students with in degree 1 \\ 
$Y$ &  Out Degree = 0 & \# of students with out degree 0 \\ 
$Y$ &  Out Degree = 1 & \# of students with out degree 1 \\
$Y$ &  Reciprocity & \# of reciprocated ties \\ 
 \hline
$X$ & Grade = 9 & \# of freshmen \\ 
$X$ & Grade = 10 & \# of sophomores \\ 
$X$ & Grade = 11 & \# of juniors \\ 
 \hline
$X,Y$ & Within Grade Homophily & Pooled homophily within grade  \\ 
$X,Y$ & +1 Grade Homophily & Pooled homophily between each grade \\ & and the grade above it \\ 
   \hline\smallskip
 \label{tbl:modterms}
\end{tabular}
\end{table}

Table \ref{tbl:supmod} shows the fitted model along with standard errors and $p-$values based upon the Fisher information matrix. We can see that students in the same grade are much more likely to be friends, as the Within Grade Homophily term is positive, and is nominally highly significant. The positive coefficient for '+1 Grade Homophily' indicates that students also tend to form connections to the grades just below or just above them.

\begin{table}
\caption{ERNM Model with Standard Errors Based on the Fisher Information}
\centering
\begin{tabular}{rrrrr}
  \hline
Term & $\hat\eta$ & Std. Error & Z & $p-$value \\ 
  \hline
Mean Degree & -217.02 & 7.81 & -27.80 & $<$0.001 \\ 
  Log Variance of degree & 25.07 & 9.06 & 2.77 & 0.006 \\ 
  In-Degree 0 & 2.62 & 0.50 & 5.20 & $<$0.001 \\ 
  In-Degree 1 & 1.05 & 0.40 & 2.62 & 0.009 \\ 
  Out-Degree 0 & 4.09 & 0.52 & 7.91 & $<$0.001 \\ 
  Out-Degree 1 & 1.93 & 0.45 & 4.25 & $<$0.001 \\ 
  Reciprocity & 2.71 & 0.23 & 11.77 & $<$0.001 \\ \hline
  Grade = 9 & 1.46 & 0.62 & 2.37 & 0.018 \\ 
  Grade = 10 & 1.93 & 0.71 & 2.72 & 0.007 \\ 
  Grade = 11 & 2.08 & 0.59 & 3.54 & $<$0.001 \\ \hline
  Grade Homophily & 4.34 & 0.46 & 9.41 & $<$0.001 \\ 
  +1 Grade Homophily & 0.63 & 0.21 & 2.98 & 0.003 \\
   \hline\smallskip
\end{tabular}
\label{tbl:supmod}
\end{table}

\begin{figure}
\centering
\includegraphics[scale=.4]{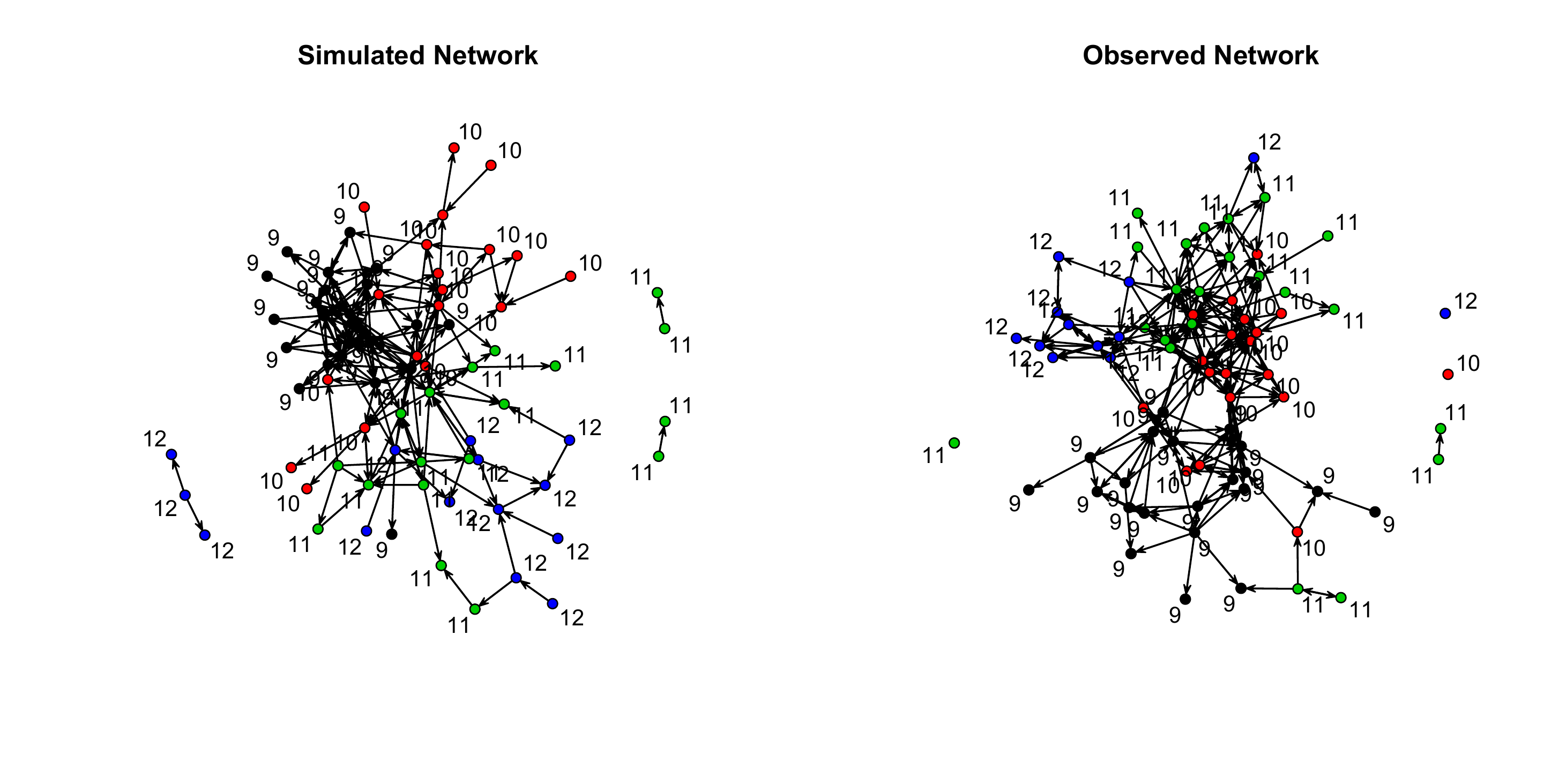}
\caption{\label{figure:simobs}
Model-Based Simulated High School }
\end{figure}

\begin{figure}
\centering
\includegraphics[scale=0.45]{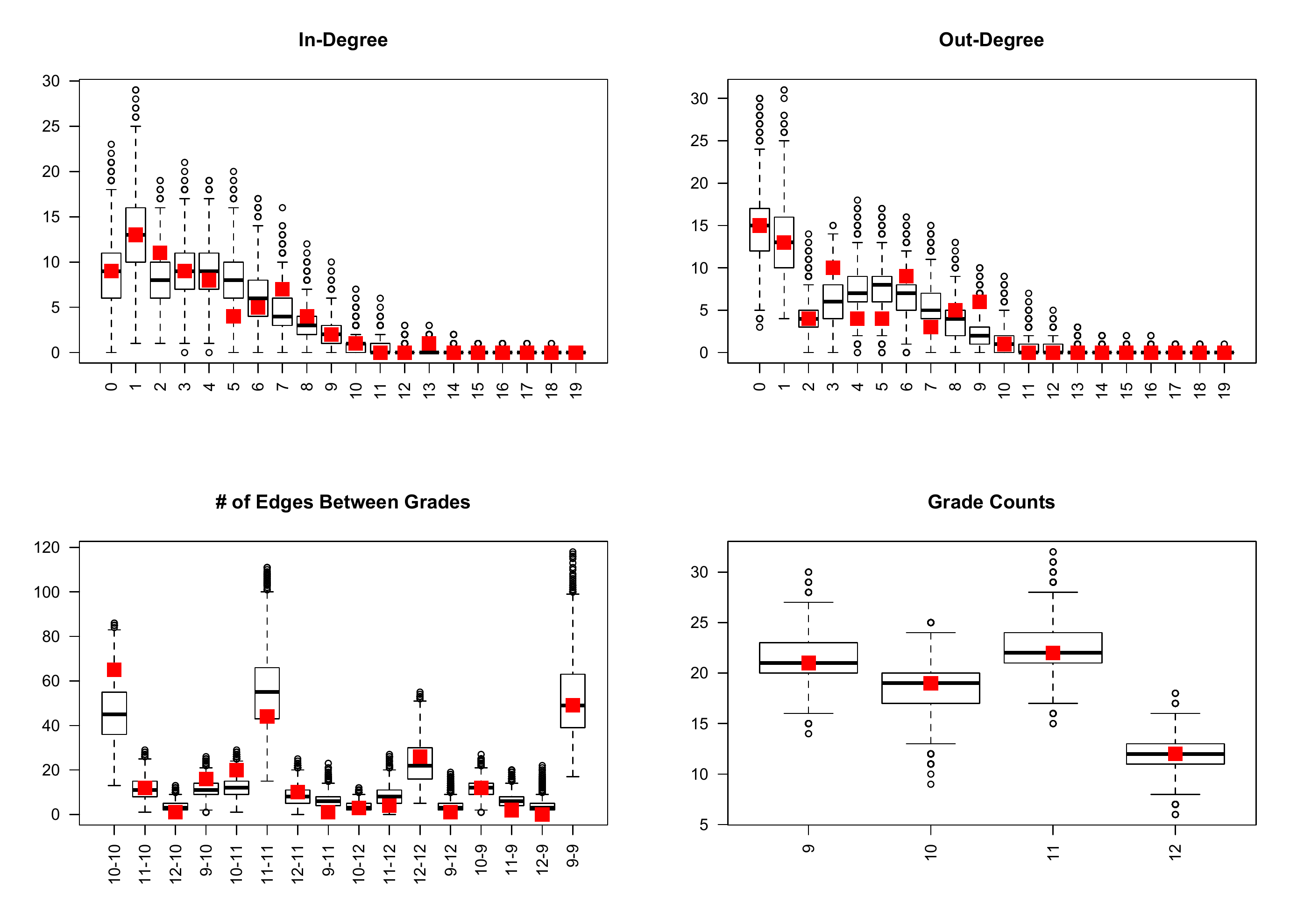}
\caption{\label{figure:stats}
Model Diagnostics}
\end{figure}

We can evaluate the fit of the model in two ways. The first is to simulate networks from the fitted model, and visually compare them to the observed network \citep{hungoodhan07}. Figure \ref{figure:simobs} shows one such simulation. The observed network and simulated network look similar, giving some support that the fitted model is reasonable. Next we can simulate network statistics from the model and compare them to the observed network. The box plots in Figure \ref{figure:stats} represent network statistics from 1000 draws from the fitted model, and the red dots are the statistics of the observed network. The degree structure matches well. Looking at the number of edges between grades, we see that the two homophily terms capture the 16 mixing statistics quite well. If desired, we could have added additional terms for each of the 16 mixing categories, but our interest was in a reasonable parsimonious representation of the network. The counts of students within each grade are perfectly centered around the observed statistics. This is expected, as these counts are explicitly included in the model, and thus the mean counts from the model match the observed counts in the high school.

\subsection{Logistic Regression on Substance Use}

One aspect of the Add Health data that is of particular interest is the degree to which students use, or have used, tobacco and alcohol. In this section we will investigate the relationship between substance use and sex. We define substance use as either current use of tobacco or having used alcohol at least 3 times. Overall 19 students reported having used substances. A naive logistic regression model with $X$ as an indicator that the sex of the adolescent is male shows a significant effect of sex (Table \ref{tbl:simplr}). Note that this model implies separability between the distribution of the network and the distribution of the outcome as in Section \ref{ss:separable}. This is an unreasonable assumption if friends tend to influence each other's substance abuse patterns, which we expect to be the case.

\begin{table}
\caption{Simple Logistic Regression Model Ignoring Network Structure. This is
the standard approach to regression in network data that ignores social
influence and selection.}
\centering
\begin{tabular}{rrrrr}
  \hline
 & $\beta$ & Std. Error & Z & $p-$value \\ 
  \hline
Intercept & -1.70 & 0.44 & -3.84 & $<$0.001 \\ 
  Gender & 1.18 & 0.57 & 2.09 & 0.037 \\ 
   \hline\smallskip
\end{tabular}
\label{tbl:simplr}
\end{table}
We extend the model in Section \ref{ss:super} with terms for substance and gender homophily, as well as terms for the logistic regression of sex on substance use. Whereas, Grade was considered random in the model in Section \ref{ss:super}, because substance use is of primary interest in this model, all covariates are fixed except for Substance use. Table \ref{tbl:jergmlr} displays the parameter estimates as well as $p$-values based on the Fisher information. Because inferences using Fisher information are typically justified using asymptotic arguments which don't apply here, we also ran a parametric bootstrap procedure with 1000 bootstraps, and bootstrap standard errors are included in Table \ref{tbl:jergmlr}. There is very close agreement between the bootstrap standard errors and the asymptotic ones, indicating that the Fisher information is a reliable measure for this model.

\begin{table}
\caption{Network Logistic Regression Parameter Estimates: These are based on the ERNM which models social influence and selection. The effect of gender on substance abuse is different than that in simple model (Table 3).}
\centering
\begin{tabular}{rrrrrr}
  \hline
  &&Bootstrap&Asymptotic&& \\
 & $\eta$&Std. Error & Std. Error & Z & $p-$value \\ 
  \hline
Mean Degree & -215.50 			& 8.32 & 8.15 & -26.44 & $<$0.001 \\ 
  Log Variance of degree & 24.46 	& 8.80 & 8.91 & 2.75 & 0.006 \\ 
  In-Degree 0 & 2.68 				& 0.55 & 0.48 & 5.55 & $<$0.001 \\ 
  In-Degree 1 & 1.07 				& 0.43 & 0.41 & 2.60 & 0.009 \\ 
  Out-Degree 0 & 4.15 			& 0.54 & 0.52 & 8.03 & $<$0.001 \\ 
  Out-Degree 1 & 1.94 			& 0.50 & 0.45 & 4.31 & $<$0.001 \\ 
  Reciprocity & 2.71 				& 0.25 & 0.23 & 11.96 & $<$0.001 \\ 
  Grade Homophily & 4.28 		& 0.44 & 0.47 & 9.18 & $<$0.001 \\ 
  +1 Grade Homophily & 0.62 		& 0.21 & 0.21 & 2.99 & 0.003 \\ \hline
  Gender Homophily & 0.78 		& 0.24 & 0.24 & 3.27 & 0.001 \\ 
  Substance Homophily & 0.76 		& 0.25 & 0.25 & 3.02 & 0.003 \\ \hline
  Intercept & -1.72 				& 0.50 & 0.44 & -3.91 & $<$0.001 \\ 
  Gender & 0.92 				& 0.55 & 0.51 & 1.79 & 0.073 \\

   \hline\smallskip
\end{tabular}
\label{tbl:jergmlr}
\end{table}

We see that the first 9 terms in the model are similar to their counterparts in Table \ref{tbl:supmod}. Two additional homophily terms are added, one for gender, and one for substance use. Both of these are highly significant, lending support to the position that it is unwise to simply perform a logistic regression ignoring network structure. The last two terms in Table \ref{tbl:jergmlr} represent the network aware logistic regression of gender of substance use, and are analogous to the terms in Table \ref{tbl:simplr}. The parameter for sex is 22\% smaller than in Table \ref{tbl:simplr} leading to a non-significant $p-$value.

\begin{figure}
\centering
\includegraphics[scale=0.65]{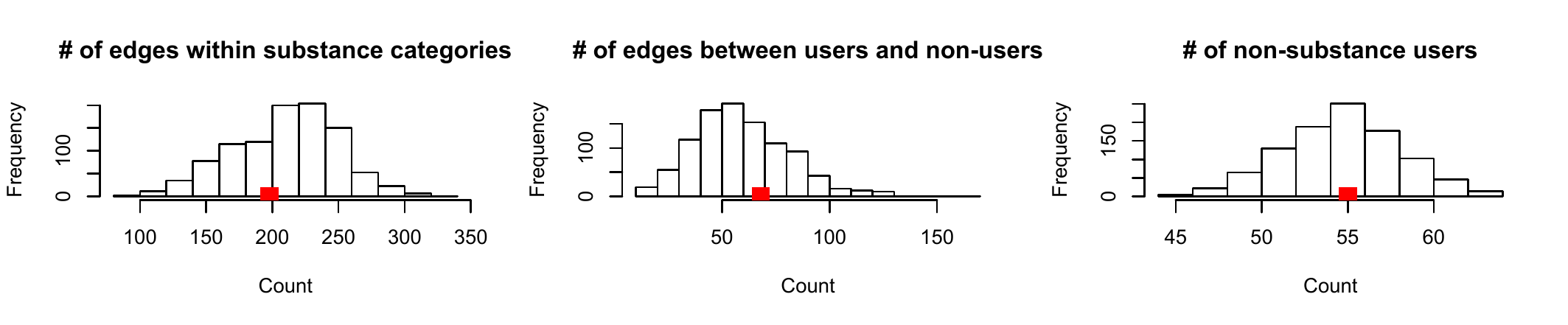}
\caption{\label{figure:subdiag}
Substance Use Homophily Diagnostics. The 
values of the observed statistics are marked in red.}
\end{figure}

Similarly to the model in Section \ref{ss:ising}, in the fitted model, 73\% of edges occur between students with the same substance abuse classification, whereas 27\% are between users and non-users. Figure \ref{figure:subdiag} shows model diagnostics for the homophily on substance abuse. Note that each marginal histogram puts high probability on the observed statistics (marked in red) and are not highly skewed, indicating that our model both captures the homophily relation, and is a reasonable model of that relation.


\section{Discussion}\label{sec:diss}

We have developed a new class of joint relational and attribute models for the analysis of network data. These models represent a generalization
of both ERGM and Gibbs random field models with each expressible as a special case of the new class. The new model provides a principled way to draw inferences about not only the graph structure, but also the nodal characteristics of the network. 

A ramification of the joint class is a natural way to specify conditional
logistic regression on nodal variables. Previous models
for network regression  have struggled with the specification due to the
ambiguity induced by endogenous nodal variable. The ERNM framework clarifies
the model formulation and the interpretation of the parameters.

Further work on specifying model statistics is necessary
to unlock the power of the ERNM class. The regularized homophily statistic of
Section 3.2 is a good illustration of the issues involves. It is a good way to
represent homophily on nodal characteristics. However, alternatives need to be
developed for other features such as transitivity.

As could be expected based on presence of degeneracy in many ERGM models, we found that there exist degenerate states in even simple ERNM models. In particular, we found that the usual statistic used to represent homophily (the major relation of interest in a joint model) displayed significant degeneracy issues, and proposed an alternative that does not.

The R package implementing the methods developed in this
paper 
will be made available
on CRAN \citep{R}.

\blind{{\relax}{
\section*{Acknowledgments}
\rm The project described was supported by grant number 1R21HD063000 from NICHD and grant number MMS-0851555 from NSF, and grant number N00014-08-1-1015 from ONR. Its contents are solely the responsibility
of the authors and do not necessarily represent the official views of the
Demographic \& Behavioral Sciences (DBS) Branch,
the National Science Foundation, or the Office of Navel Research.
The authors would like to thank the members of the Hard-to-Reach Population Research Group (\url{hpmrg.org}), especially Krista J. Gile, for their helpful input.
This research uses data from Add Health, a program project designed by J.
Richard Udry, Peter S. Bearman, and Kathleen Mullan Harris, and funded by Grant
P01-HD31921 from the Eunice Kennedy Shriver National Institute of Child Health
and Human Development, with cooperative funding from 17 other agencies. Special
acknowledgment is due Ronald R. Rindfuss and Barbara Entwisle for assistance in
the original design. Persons interested in obtaining data files from Add Health
should contact Add Health, Carolina Population Center, 123 W. Franklin Street,
Chapel Hill, NC 27516-2524 (addhealth@unc.edu). No direct support was received
from Grant P01-HD31921 for this analysis.
}}

\app{
\appendix
\section*{Appendix A: Specifics of ERNM Terms}\label{app:ernm-terms}
\gdef\thesection{A}
Here we explicitly define the network terms in \eqref{tbl:modterms}. Let $n$ be then number of nodes in the network, $d_{i,j}^x = \sum_ky_{i,k}I(x_k=j) + \sum_ky_{k,i}I(x_k=j)$ be the degree of node $i$ to category $j$ of $x$, and $d_i^+ = \sum_ky_{k,i}$, $d_i^-\sum_ky_{i,k}$, $d_i=d_i^+ + d_i^-$ be the in, out and overall degree respectively. Then the model terms can be expressed as:
\begin{eqnarray}
{\rm mean\ degree} &=& \frac{\sum_i^nd_i}{n} \nonumber \\
{\rm log\ variance\ of\ degree} &=& {\rm log}(\frac{\sum_i^n(mean\ degree - d_i)^2}{n}) \nonumber \\
{\rm in degree}\ k &=& \sum_i^nI(d_i^-=k)  \nonumber \\
{\rm out degree}\ k &=&\sum_i^nI(d_i^+=k) \nonumber \\
{\rm reciprocity}\  &=&\sum_i^n \sum_j^n y_{i,j}y_{j,i} \nonumber \\
{\rm within\ grade\ homophily} &=& \sum_{k\in \{9,10,11,12\}}\sum_{i:grade=k} \sqrt{d_{i,k}} - E_{\Perp}(\sqrt{d_{i,k}}) \nonumber \\
{\rm +1\ grade\ homophily} &=& \sum_{k\in \{9,10,11\}}\sum_{i:grade=k} \sqrt{d_{i,k+1}} - E_{\Perp}(\sqrt{d_{i,k+1}}) + \nonumber \\
 && \sum_{k\in \{10,11,12\}}\sum_{i:grade=k} \sqrt{d_{i,k-1}} - E_{\Perp}(\sqrt{d_{i,k-1}}) \nonumber 
\end{eqnarray}

For large networks some computational efficiency can be obtained by
approximating the the expectations $E_{\Perp}(\sqrt{d_{i,k}})$ by that of
the square root of a binomial variable, with probability equal to the proportion of nodes in category $l$, and size equal to the out-degree of node $i$. Each term of the sum is then the square root of the number of connections to category $l$, from node $i$, minus what would be expected by chance. Note that the expectation would more accurately be a hypergeometric distribution, due to the fact that only one edge can connect two nodes, however, the binomial approximation is much faster to compute and is asymptotically correct for sparse graphs.
This approach was used in the application of Section \ref{sec:app}.

\section*{Appendix B: An MCMC algorithm for ERNM}\label{app:ernm-algorithm}
\gdef\thesection{B}

We use a Metropolis-Hastings algorithm to sample from an ERNM \citep{gil96}. The algorithm alternates between proposing a change to a dyad with probability $p_{dyad}$ and proposing a change to a nodal variable. Because the graphs for social networks are usually sparse, when proposing a dyad change the algorithm selects an edge to remove with probability $p_{edge}$ and a random dyad to toggle with probability $1-p_{edge}$. We found that this leads to better mixing than simply toggling a random dyad \citep{ergmtermsjss}. When proposing a change to the nodal attributes, an attribute is picked at random. If it is categorical, a random new category is chosen. If it is continuous, it is perturbed by adding a small constant $\epsilon$.

\def\RandomEdge{{\rm RandomEdge}}
\def\RandomDyad{{\rm RandomDyad}}
\def\NumberOfEdges{{\rm NumberOfEdges}}
\def\NumberOfDyads{{\rm NumberOfDyads}}
\def\RandomEdge{{\rm RandomEdge}}
\def\Uniform{{\rm Uniform}}
\def\RandomAttribute{{\rm RandomAttribute}}
\def\RandomCategory{{\rm RandomCategory}}
\def\HasEdge{{\rm HasEdge}}
\def\IsContinuous{{\rm IsContinuous}}
\def\Normal{{\rm Normal}}
The following algorithm can be used to generate a random draw from an 
ERNM probability distribution~\eqref{eq:lik}
with an intractable normalizing constant:
\advance\baselineskip by -3pt
\begin{algorithmic}[1]
  \REQUIRE Arbitrary $(y^{0}, x^0) \in nets(Y,X)$, $p_{dyad}\in [0,1]$, $p_{edge}\in [0,1]$ and $S$ sufficiently large
  \FOR{$s\gets 1$ to $S$}
  \STATE $y^* \gets y^{(s-1)}$
  \STATE $x^* \gets x^{(s-1)}$
  \STATE $u_{dyad} \gets \Uniform(0,1)$
  \IF{$u_{dyad}<p_{dyad}$}
  	 \STATE $u_{edge} \gets \Uniform(0,1)$
	 \IF{$u_{edge} < p_{edge}$}
	 	\STATE $(i,j) \gets \RandomEdge(y^*)$
		\STATE $y_{i,j}^* \gets 0$
		\STATE $q \gets \frac{\NumberOfEdges(y*)}{\NumberOfEdges(y*) + \NumberOfDyads(y*)}$
	\ELSE
		\STATE $(i,j) \gets \RandomDyad(y^*)$
		\IF{$y_{i,j}^*=0$}
			\STATE $y_{i,j}^* \gets 1$
			\STATE $q \gets \frac{\NumberOfEdges(y*)}{\NumberOfEdges(y*) + \NumberOfDyads(y*)}$
		\ELSE
			\STATE $y_{i,j}^* \gets 0$
			\STATE $q \gets 1 + \frac{\NumberOfDyads(y*)}{\NumberOfEdges(y*) + 1}$
		\ENDIF	
	 \ENDIF
 \ELSE
 	\STATE $(k,l) \gets \RandomAttribute(x^*)$
	\IF{$\IsContinuous(x^*_{*,l})$}
		\STATE $\epsilon \gets \Normal(0,\sigma)$
		\STATE $x_{k,l}^* \gets x_{k,l}^* + \epsilon$ 
		\STATE $q \gets 1$
	\ELSE
		\STATE $x_{k,l}^* \gets \RandomCategory(x_{*,l}^*)$ 
		\STATE $q \gets 1$		
	\ENDIF
  \ENDIF
  \STATE $r \gets q e^{\eta(g(x^*,y^*) -g(x^{(s-1)},y^{(s-1)}))}$
  \STATE $u \gets \Uniform(0,1)$
  \IF{$ u<r $}
  \STATE $(y^{s},x^{s}) \gets (y^{*},x^{*})$
  \ELSE
  \STATE $(y^{s},x^{s}) \gets (y^{s-1},x^{s-1})$
  \ENDIF
  \ENDFOR
  \RETURN $(y^{S},x^{S})$
\end{algorithmic}
\advance\baselineskip by 3pt

Note that an adjustment to the calculation of $q$ must be made when toggling the graph when less than two edges are present in the network. If we are removing the last edge, then $q \gets 1/(\NumberOfDyads(y*) + .5)$, and if we are adding an edge to an empty graph, then $q \gets 0.5(\NumberOfDyads(y*) + 1)$.

In order for this algorithm to be fast, we must calculate the likelihood ratio\newline
$e^{\eta{\cdot}(g(x^*,y^*) -g(x^{(s-1)},y^{(s-1)}))}$ quickly, preferably in constant time relative to the size of the network. We do this with change statistics \citep{ergmtermsjss}, which can quickly calculate the differences in the $h$ statistics given small changes to the graph $y$ or nodal attributes $x$.
\citet{ergmtermsjss} review change statistics for commonly used ERGM terms and
these can be reused here for changes in the graph (i.e. $g(x^{(s-1)},y^*) - g(x^{(s-1)},y^{(s-1)})$). ERNM require additional terms, such as those
specified in Section \ref{ss:homo}, and also require that all change statistics be generalized to allow for changes in nodal attributes (i.e. $g(x^*,y^{(s-1)}) - g(x^{(s-1)},y^{(s-1)})$).
}

\addcontentsline{toc}{section}{References}
\bibliographystyle{jrss}
\bibliography{jrssbernm.bib}

\end{document}